\documentclass[final,5p,times,twocolumn]{elsarticle}
\usepackage{graphicx}
\usepackage{amssymb}
\usepackage{amsmath}
\usepackage[margin=0pt,font=small,labelfont=bf,labelsep=colon]{caption}
\usepackage{algorithm}
\usepackage{algpseudocode}
\usepackage{listings}
\usepackage{color}
\usepackage{textcomp}
\usepackage{url}
\usepackage{lineno}
\newcommand{\vect}[1]{\boldsymbol{#1}}
\biboptions{sort&compress}
\journal{Computer Physics Communications}

\begin{document}

\begin{frontmatter}

\title{Efficient molecular dynamics simulations with many-body potentials on graphics processing units}

\author{Zheyong Fan $^{1,2}$ \corref{cor1}}
\ead{brucenju@gmail.com}
\author{Wei Chen$^{3}$ \corref{cor1} }
\ead{weichen@cnic.cn}
\author{Ville Vierimaa $^{2}$}
\author{Ari Harju $^{2}$}
\cortext[cor1]{Corresponding authors}
\address{$^{1}$School of Mathematics and Physics, Bohai University, Jinzhou, China}
\address{$^{2}$COMP Centre of Excellence and Helsinki Institute of Physics,
Department of Applied Physics, Aalto University, Helsinki, Finland}
\address{$^{3}$Computer Network Information Center, Chinese Academy of Sciences, P.O. Box 349, 100190 Beijing, China}

\begin{abstract}
Graphics processing units have been extensively used to accelerate classical molecular dynamics simulations. However, there is much less progress on the acceleration of force evaluations for many-body potentials compared to pairwise ones. In the conventional force evaluation algorithm for many-body potentials, the force, virial stress, and heat current for a given atom are accumulated within different loops, which could result in write conflict between different threads in a CUDA kernel. In this work, we provide a new force evaluation algorithm, which is based on an explicit pairwise force expression for many-body potentials derived recently [Phys. Rev. B 92 (2015) 094301]. In our algorithm, the force, virial stress, and heat current for a given atom can be accumulated within a single thread and is free of write conflicts. We discuss the formulations and algorithms and evaluate their performance. A new open-source code, GPUMD, is developed based on the proposed formulations. For the Tersoff many-body potential, the double precision performance of GPUMD using a Tesla K40 card is equivalent to that of the LAMMPS (Large-scale Atomic/Molecular Massively Parallel Simulator) molecular dynamics code running with about 100 CPU cores (Intel Xeon CPU X5670 @ 2.93 GHz).
\end{abstract}

\begin{keyword}
Molecular dynamics simulation \sep
Many-body potential \sep
Tersoff potential \sep
Stillinger-Weber potential \sep
Graphics processing units \sep
Virial stress \sep
Heat current
\end{keyword}

\end{frontmatter}

\section{Introduction}
\label{Introduction}

Molecular dynamics (MD) simulation is one of the most important numerical tools in investigating various physical properties of materials. Many applications using MD simulation demand high performance computing. In the past decade, the computational power of general-purpose graphics processing units (GPUs) has been exploited to accelerate many MD simulations. Not only existing MD codes and libraries, such as AMBER \cite{gotz2012}, Gromacs \cite{pall2013,kutzner2015}, LAMMPS \cite{plimpton1995,brown2011,brown2012}, NAMD \cite{stone2007}, and OpenMM \cite{eastman2010} have been benefited from utilizing GPUs as accelerators, but also new codes, such as HOOMD-blue \cite{anderson2008,anderson2013,glaser2015}, HALMD \cite{colberg2011}, and RUMD \cite{bailey2015}, have been built from the ground up to achieve high performance using one or more GPUs.

Most of the previous relevant works have only considered pairwise potentials, or a special many-body potential, namely, the embedded atom method \cite{morozov2011,brown2012b,hou_qing2013}, which are relatively simple to implement on GPUs. GPU-acceleration of many-body potentials such as the Tersoff \cite{tersoff1989}, Stillinger-Weber \cite{stillinger1985}, and Brenner \cite{brenner2002} potentials, which play an important role in modelling various materials, is more challenging and has only attracted some attention recently \cite{hou2013,brown2013,knizhnik2015,hohnerbach2016,tredak2016,nguyen2017}. Taking three-body interaction as an example, a naive implementation of the force evaluation function, as usually done in a serial CPU code, requires accumulating the forces on three different atoms within a single thread. In a GPU kernel with many threads, each atom is usually associated with one thread and the force accumulation for an atom from the thread it belongs to will conflict with the force accumulation for the same atom from another thread. This causes a problem called write conflict where two threads try to write data simultaneously into the same global memory \cite{cuda}. One way to avoid write conflict is to use atomic operations, which are usually quite slow and can also introduce randomness in the computation, which is undesirable for debugging.

There have been some proposals to avoid using atomic operations. Hou \textit{et al.} \cite{hou2013} proposed an algorithm for implementing the Tersoff potential on a GPU, which has achieved impressive performance, but requires using a special fixed neighbour list and is thus not quite flexible. Brown and Yamada \cite{brown2013} proposed a flexible GPU-implementation of the Stillinger-Weber potential within LAMMPS, which is free of write conflicts. A similar proposal was given by Knizhnik \textit{et al.} \cite{knizhnik2015}. Recently, H\"{o}hnerbach \textit{et al.} \cite{hohnerbach2016} developed a vectorization scheme to achieve performance portability across various parallel computing platforms for the Tersoff potential within LAMMPS. GPU-acceleration of the more complicated second-generation REBO potential \cite{brenner2002} has also been studied by Tredak \textit{et al.} \cite{tredak2016}. In a recently published work \cite{nguyen2017} (after we submitted this paper), Nguyen reported significant speedups for MD simulations with Tersoff-type potentials using one or more high-end GPUs.

Here, we propose a general algorithm of force evaluation for many-body potentials and present details of its GPU-implementation and performance. The new force evaluation algorithm is based on an explicit pairwise force expression for many-body potentials derived recently \cite{fan2015}. In this approach, the force, virial stress, and heat current for a given atom are well defined and can be accumulated within a single thread. Therefore, write conflict is absent by construction. To be specific, we discuss the algorithm explicitly in terms of the Tersoff potential, but performance evaluation is made for both the Tersoff potential and the Stillinger-Weber potential. The implementation is done based on a previous work \cite{fan2013}, and the resulting code, which we call GPUMD (Graphics Processing Units Molecular Dynamics), will be made public soon. Using silicon crystal as a test system, we measure the performance of GPUMD and compare it with LAMMPS.

\section{Formulations and algorithms}
\label{section:Theory}

\subsection{The Tersoff many-body potential}

Although the method to be introduced is applicable to any many-body potential, it is beneficial to start with an explicit example, which is taken as the widely used Tersoff potential. Generalizations to other many-body potentials will be discussed later.

The total potential energy for a system of $N$ atoms described by the Tersoff potential can be written as \cite{tersoff1989}
\begin{equation}
U = \frac{1}{2}\sum_i\sum_{j\neq i} U_{ij},
\end{equation}
where
\begin{equation}
\label{equation:Tersoff_U_ij}
U_{ij} = f_C(r_{ij}) \left( f_R(r_{ij}) - b_{ij} f_A(r_{ij}) \right),
\end{equation}
\begin{equation}
b_{ij} = \left(1 + \beta^n \zeta^n_{ij}\right)^{-\frac{1}{2n}},
\end{equation}
\begin{equation}
\zeta_{ij} = \sum_{k\neq i, j}f_C(r_{ik}) g_{ijk},
\end{equation}
\begin{equation}
g_{ijk} = 1 + \frac{c^2}{d^2} - \frac{c^2}{d^2+(h-\cos\theta_{ijk})^2}.
\end{equation}
Here, $\beta$, $n$, $c$, $d$, and $h$ are material-specific parameters and $\theta_{ijk}$ is the angle formed by $\vect{r}_{ij}$ and $\vect{r}_{ik}$, which implies that
\begin{equation}
\label{equation:cos_theta}
\cos\theta_{ijk} = \cos\theta_{ikj} = \frac{\vect{r}_{ij} \cdot \vect{r}_{ik}}{r_{ij} r_{ik}}.
\end{equation}
Our convention is that $\vect{r}_{ij} \equiv \vect{r}_j - \vect{r}_i$ represents the position difference pointing from atom $i$ to atom $j$. The magnitude of $\vect{r}_{ij}$ is denoted as $r_{ij}$.

As in many empirical potentials, the energy $U_{ij}$ consists of a repulsive part $ f_R(r_{ij})$ and an attractive part $-b_{ij}  f_A(r_{ij})$. The many-body nature of the Tersoff potential is embodied in the bond order function $b_{ij}$ appearing in the attractive part, the value of which depends not only on $\vect{r}_i$ and $\vect{r}_j$, but also on the positions of other atoms near atom $i$.

The function $f_C(r_{ij})$ is a cutoff function, which is only nonzero when $r_{ij}$ is less than a cutoff distance. Therefore, a Verlet neighbour list can be used to speed up the force evaluation. For uniform cutoff, the standard cell list method is very efficient, although more sophisticated methods perform better for systems with large size disparities \cite{howard2016}.

For simplicity, we have presented the original Tersoff potential formulation in a form suitable for single-element systems. Our algorithm and implementation are more general, which can treat systems with more than one type of atom or systems described by a modified formulation of the Tersoff potential.

\subsection{The conventional method of implementing the Tersoff potential}

Due to the three-body nature of the Tersoff potential, the conventional method for evaluating the interatomic forces is significantly different from that in the case of a simple two-body potential. Algorithm \ref{algorithm:conventional_algorithm} presents a pseudo code for the conventional method as implemented in most existing MD codes such as LAMMPS \cite{plimpton1995}. The following symbols are used:
\begin{itemize}
\item $N$: number of atoms
\item $U_i$: potential energy of atom $i$
\item $\vect{F}_i$: total force on atom $i$
\item $\textbf{W}_i$: per-atom virial stress of atom $i$
\item NN$_i$: number of neighbour atoms of atom $i$
\item NL$_{im}$: index of the $m$th neighbour atom of atom $i$
\item $\vect{J}_i$: per-atom heat current of atom $i$
\end{itemize}

\begin{algorithm}[htb]
\caption{Pseudo code for the conventional method of evaluating many-body force and related quantities.}
\label{algorithm:conventional_algorithm}
\begin{algorithmic}[1]
\For {$i$ = 0 to $N - 1$}
    \State Initialise $U_i$, $\vect{F}_i$, and $\textbf{W}_i$ to zero
\EndFor
\For {$i$ = 0 to $N - 1$}
    \For {$m$ = 0 to NN$_i - 1$}
        \State $j \leftarrow \text{NL}_{im}$
        \State $U_i \leftarrow U_i + \frac{1}{2}U_{ij}$
        \State $\vect{F}_i \leftarrow \vect{F}_i + \vect{F}^{(ij)}_{i}$
        \State $\vect{F}_j \leftarrow \vect{F}_j + \vect{F}^{(ij)}_{j}$
        \State $\textbf{W}_i \leftarrow \textbf{W}_i - \frac{1}{2} \vect{r}_{ij} \otimes \vect{F}^{(ij)}_{i}$
        \For {$n$ = 0 to NN$_i - 1$}
            \State $k \leftarrow \text{NL}_{in}$
            \If {$k = j$}
                \State Continue
            \EndIf
            \State $\vect{F}_i \leftarrow \vect{F}_i + \vect{F}^{(ijk)}_{i}$
            \State $\vect{F}_j \leftarrow \vect{F}_j + \vect{F}^{(ijk)}_{j}$
            \State $\vect{F}_k \leftarrow \vect{F}_k + \vect{F}^{(ijk)}_{k}$
            \State $\textbf{W}_i \leftarrow \textbf{W}_i + \frac{1}{3}
                   \left(\vect{r}_{ij} \otimes \vect{F}^{(ijk)}_{j} + \vect{r}_{ik} \otimes \vect{F}^{(ijk)}_{k}\right)$
            \State $\textbf{W}_j \leftarrow \textbf{W}_j + \frac{1}{3}
                   \left(\vect{r}_{ij} \otimes \vect{F}^{(ijk)}_{j} + \vect{r}_{ik} \otimes \vect{F}^{(ijk)}_{k}\right)$
            \State $\textbf{W}_k \leftarrow \textbf{W}_k + \frac{1}{3}
                   \left(\vect{r}_{ij} \otimes \vect{F}^{(ijk)}_{j} + \vect{r}_{ik} \otimes \vect{F}^{(ijk)}_{k}\right)$
        \EndFor
    \EndFor
\EndFor
\For {$i$ = 0 to $N - 1$}
    \State $\vect{J}_i \leftarrow \textbf{W}_i \cdot \vect{v}_i$
\EndFor
 \end{algorithmic}
\end{algorithm}

In Algorithm \ref{algorithm:conventional_algorithm}, the potential energy $U_i \equiv  \sum_{j \neq i} U_{ij}/2$ is  accumulated in line 7, the two-body parts of the force and per-atom virial stress are accumulated in lines 8-9 and 10, respectively, and the many-body parts of the force and per-atom virial stress are accumulated in lines 16-18 and 19-21, respectively. Last, in line 26, the per-atom heat current is calculated from the per-atom virial stress and velocity.

The forces defined in the pseudo code can be explicitly written as
\begin{equation}
\label{equation:F_ij_i}
  \vect{F}^{(ij)}_{i} =
      - \frac{1}{2} \frac{\partial}{\partial \vect{r}_i} \left(f_C(r_{ij}) f_R(r_{ij})\right)
      + \frac{1}{2}b_{ij} \frac{\partial}{\partial \vect{r}_i} \left(f_C(r_{ij}) f_A(r_{ij})\right),
\end{equation}
\begin{equation}
\label{equation:F_ij_j}
  \vect{F}^{(ij)}_{j} =
      - \frac{1}{2} \frac{\partial}{\partial \vect{r}_j} \left(f_C(r_{ij}) f_R(r_{ij})\right)
      + \frac{1}{2}b_{ij} \frac{\partial}{\partial \vect{r}_j} \left(f_C(r_{ij}) f_A(r_{ij})\right),
\end{equation}
\begin{equation}
\label{equation:F_ijk_i}
   \vect{F}^{(ijk)}_{i} = \frac{1}{2} f_C(r_{ij})f_A(r_{ij})\frac{\partial b_{ij}}{\partial \zeta_{ij}}
   \frac{\partial}{\partial \vect{r}_i} \left(f_C(r_{ik})g_{ijk}\right),
\end{equation}
\begin{equation}
\label{equation:F_ijk_j}
   \vect{F}^{(ijk)}_{j} = \frac{1}{2} f_C(r_{ij})f_A(r_{ij})\frac{\partial b_{ij}}{\partial \zeta_{ij}}
   \frac{\partial}{\partial \vect{r}_j} \left(f_C(r_{ik})g_{ijk}\right),
\end{equation}

\begin{equation}
\label{equation:F_ijk_k}
   \vect{F}^{(ijk)}_{k} = \frac{1}{2} f_C(r_{ij})f_A(r_{ij})\frac{\partial b_{ij}}{\partial \zeta_{ij}}
   \frac{\partial}{\partial \vect{r}_k} \left(f_C(r_{ik})g_{ijk}\right).
\end{equation}

Here, only the repulsive part of $\vect{F}^{(ij)}_{i}$ [the first term on the right hand side of Eq. (\ref{equation:F_ij_i})] is pairwise; all the other force expressions are not. Newton's third law could be exploited to reduce calculations regarding the pairwise part, but this would not result in a noticeable improvement of the overall performance, because the pairwise part only takes up a tiny fraction of the whole force evaluation. Therefore, Newton's third law has not been used in Algorithm \ref{algorithm:conventional_algorithm}. Regardless of using Newton's third law for the pairwise part or not, the above conventional method can be straightforwardly implemented on the CPU.

Newton's third law has also not been used on the GPU implementation \cite{anderson2008} of two-body potentials due to the problem of concurrent writes (write conflict) to the same location in global memory. When two or more threads in the same warp write to the same location in global memory, only one thread performs the write and it is not defined which thread does it \cite{cuda}. To understand why this feature of CUDA restricts the use of Newton's third law on the GPU, we recall that in the commonly used force-evaluation kernel \cite{anderson2008} for two-body potentials, which we call the thread-scheme \cite{fan2013}, one thread is devoted to the calculation of the total force on one atom. Using Newton's third law would require accumulating the total force on atom $i$ by two threads. If these two threads are in the same warp, the total force on atom $i$ would not be correctly accumulated. One may try to use atomic operations to solve this problem, but this would hardly result in a gain of performance.

In view of the above discussion, one would immediately realise the difficulty of implementing Algorithm \ref{algorithm:conventional_algorithm} on the GPU: the partial accumulations of the forces and stresses on atoms $j$ and $k$ by the thread associated with atom $i$ (cf. lines 9, 17-18, 20-21 in Algorithm \ref{algorithm:conventional_algorithm}) would conflict with those by the threads associated with atoms $j$ and $k$. Therefore, Algorithm \ref{algorithm:conventional_algorithm} is not suitable for GPU-implementation. This difficulty has also been realised previously and some strategies are proposed to circumvent it \cite{hou2013,brown2013}. Below, we present a general algorithm for many-body potentials which can lead to efficient GPU-implementation. 

\subsection{A new method of implementing the Tersoff potential}

Despite the many-body nature of the Tersoff potential, a pairwise force expression that complies with Newton's third law has been derived recently \cite{fan2015}:
\begin{equation}
\label{equation:F_i_New}
\vect{F}_i = \sum_{j \neq i} \vect{F}^{\text{Tersoff}}_{ij},
\end{equation}
\begin{equation}
\label{equation:F_ij_New}
\vect{F}^{\text{Tersoff}}_{ij} = - \vect{F}^{\text{Tersoff}}_{ji}
= \frac{1}{2}\frac{\partial}{\partial \vect{r}_{ij}}
\left( U_{ij} + U_{ji} + \sum_{k\neq i, j} \left(U_{ik} + U_{jk}\right) \right),
\end{equation}
which can be simplified in terms of the per-atom potential to be
\begin{equation}
\label{equation:F_ij_simplified}
\vect{F}^{\text{Tersoff}}_{ij} =
    \frac{\partial U_i}{\partial \vect{r}_{ij}} -
    \frac{\partial U_j}{\partial \vect{r}_{ji}}.
\end{equation}
An explicit expression for $\partial U_i/\partial \vect{r}_{ij}$ can be found in Ref. \cite{fan2015}.

One may wonder whether the new pairwise force expression $\vect{F}^{\text{Tersoff}}_{ij}$ produces per-atom forces $\vect{F}_i$ that are equivalent to those obtained by the conventional method. The answer should be definitely yes; otherwise, either the conventional or our new method of force evaluation is wrong. To show this equivalence, let us note that according to Algorithm \ref{algorithm:conventional_algorithm}, the total force acting on atom $i$ can be written as:
\begin{equation}
  \vect{F}_i = \sum_{j \neq i} \left(\vect{F}^{(ij)}_{i} + \vect{F}^{(ji)}_{i}\right)
             + \sum_{j \neq i} \sum_{k \neq i, j}
               \left(
               \vect{F}^{(ijk)}_{i} + \vect{F}^{(jik)}_{i} + \vect{F}^{(jki)}_{i}
               \right).
\end{equation}
Plugging in Eqs. (\ref{equation:F_ij_i})-(\ref{equation:F_ijk_k}), and changing the absolute positions $\vect{r}_i$ to relative ones $\vect{r}_{ij}$ using the chain rule, we get
\begin{align}
\label{equation:F_i_LAMMPS}
  \vect{F}_i
  &= \frac{1}{2} \sum_{j \neq i} \frac{\partial}{\partial \vect{r}_{ij}} \left(f_C(r_{ij}) f_R(r_{ij})\right)
   - \frac{1}{2} \sum_{j \neq i} b_{ij} \frac{\partial}{\partial \vect{r}_{ij}}
                 \left(f_C(r_{ij}) f_A(r_{ij})\right) \nonumber \\
  &+ \frac{1}{2} \sum_{j \neq i} \frac{\partial}{\partial \vect{r}_{ij}} \left(f_C(r_{ji}) f_R(r_{ji})\right)
   - \frac{1}{2} \sum_{j \neq i} b_{ji} \frac{\partial}{\partial \vect{r}_{ij}}
                 \left(f_C(r_{ji}) f_A(r_{ji})\right) \nonumber \\
  &- \frac{1}{2} \sum_{j \neq i} \sum_{k \neq i, j} f_C(r_{ij}) f_A(r_{ij}) \frac{\partial b_{ij}}{\partial \zeta_{ij}}
                 \frac{\partial}{\partial \vect{r}_{ij}} \left(f_C(r_{ik})g_{ijk}\right) \nonumber \\
  &- \frac{1}{2} \sum_{j \neq i} \sum_{k \neq i, j} f_C(r_{ij}) f_A(r_{ij}) \frac{\partial b_{ij}}{\partial \zeta_{ij}}
                 \frac{\partial}{\partial \vect{r}_{ik}} \left(f_C(r_{ik})g_{ijk}\right) \nonumber \\
  &- \frac{1}{2} \sum_{j \neq i} \sum_{k \neq i, j} f_C(r_{ji}) f_A(r_{ji}) \frac{\partial b_{ji}}{\partial \zeta_{ji}}
                 \frac{\partial}{\partial \vect{r}_{ij}} \left(f_C(r_{jk})g_{jik}\right) \nonumber \\
  &- \frac{1}{2} \sum_{j \neq i} \sum_{k \neq i, j} f_C(r_{jk}) f_A(r_{jk}) \frac{\partial b_{jk}}{\partial \zeta_{jk}}
                 \frac{\partial}{\partial \vect{r}_{ij}} \left(f_C(r_{ji})g_{jki}\right).
\end{align}
In this equation, the sum of the 1st and 4th lines is $1/2\sum_{j\neq i} \partial U_{ij} / \partial \vect{r}_{ij} $, the sum of the 2nd and 5th lines is $1/2 \sum_{j\neq i} \partial U_{ji} / \partial \vect{r}_{ij}$, the 3rd line, with the dummy indices $j$ and $k$ interchanged, is $1/2 \sum_{j\neq i} \sum_{k \neq i, j} \partial U_{ik} / \partial \vect{r}_{ij} $, and the 6th line is $1/2 \sum_{j\neq i} \sum_{k \neq i, j} \partial U_{jk} / \partial \vect{r}_{ij} $. Therefore, the conventional expression Eq.~(\ref{equation:F_i_LAMMPS}) is equivalent to Eqs.~(\ref{equation:F_i_New}) and (\ref{equation:F_ij_New}).

While there is no difference between the force calculations based on the conventional and the new methods, the same cannot be said for some other quantities such as the virial stress and heat current. The total virial stress tensor is defined as
\begin{equation}
\textbf{W} = \sum_i \vect{r}_i \otimes \vect{F}_i.
\end{equation}
In MD simulations with periodic boundary conditions in one or more directions, the absolute positions cause problems and in the case of two-body potentials, one can change them to relative positions using Newton's third law. The resulting virial stress tensor takes a simple form:
\begin{equation}
\textbf{W} = - \frac{1}{2} \sum_i \sum_{j\neq i} \vect{r}_{ij} \otimes \vect{F}_{ij}.
\end{equation}
Here, $\vect{F}_{ij} = - \vect{F}_{ji}$ is the pairwise force acting on atom $i$ by atom $j$. One can also decompose the total virial stress into per-atom ones:
\begin{equation}
\textbf{W} = \sum_i \textbf{W}_i,
\end{equation}
\begin{equation}
\textbf{W}_i = - \frac{1}{2} \sum_{j\neq i} \vect{r}_{ij} \otimes \vect{F}_{ij}.
\end{equation}
Since pairwise forces $\vect{F}^{\text{Tersoff}}_{ij}$ also exist for the Tersoff potential, the per-atom virial stress tensor for the Tersoff potential takes the same simple form as in the case of two-body potentials:
\begin{equation}
\label{equation:W_i_Tersoff}
\textbf{W}^{\text{Tersoff}}_i
= - \frac{1}{2} \sum_{j\neq i} \vect{r}_{ij} \otimes \vect{F}^{\text{Tersoff}}_{ij}.
\end{equation}
However, the existence of a pairwise force expression has not been widely recognised and the virial stress tensor in standard MD packages such as LAMMPS is not implemented in this way. Referring to Algorithm \ref{algorithm:conventional_algorithm}, the per-atom virial stress tensor as implemented in LAMMPS takes a rather complicated form:
\begin{align}
\textbf{W}_i = &-\frac{1}{2}\sum_{j \neq i} \vect{r}_{ij} \otimes \vect{F}^{(ij)}_{i} \nonumber \\
               &+ \frac{1}{3} \sum_{j \neq i} \sum_{k \neq i, j}
               \left(
                   \vect{r}_{ij} \otimes \vect{F}^{(ijk)}_{j} + \vect{r}_{ik} \otimes \vect{F}^{(ijk)}_{k}
               \right) \nonumber \\
               &+ \frac{1}{3} \sum_{j \neq i} \sum_{k \neq i, j}
               \left(
                   \vect{r}_{ji} \otimes \vect{F}^{(jik)}_{i} + \vect{r}_{jk} \otimes \vect{F}^{(jik)}_{k}
               \right) \nonumber \\
               &+ \frac{1}{3} \sum_{j \neq i} \sum_{k \neq i, j}
               \left(
                   \vect{r}_{kj} \otimes \vect{F}^{(kji)}_{j} + \vect{r}_{ki} \otimes \vect{F}^{(kji)}_{i}
               \right).
\end{align}
This expression is not likely equivalent to Eq.~(\ref{equation:W_i_Tersoff}). For example, the first line in this equation suggests that the force component $\vect{F}^{(ij)}_{i}$ has been taken to be pairwise, which is not true because $b_{ij} \neq b_{ji}$.

A related quantity is the potential part of the heat current $\vect{J}$. For two-body potentials, it can be written as
\begin{equation}
\label{equation:j_pot_pair}
  \vect{J}
  = - \frac{1}{2}\sum_i \sum_{j \neq i}
    \vect{r}_{ij} (\vect{F}_{ij} \cdot \vect{v}_i).
\end{equation}
Since $\vect{r}_{ij} (\vect{F}_{ij} \cdot \vect{v}_i) = (\vect{r}_{ij} \otimes \vect{F}_{ij}) \cdot \vect{v}_i$, we can also express $\vect{J}$ in terms of the per-atom virial stress tensor:
\begin{equation}
\label{equation:j_pot_pair_stress}
  \vect{J} = \sum_{i} \textbf{W}_{i} \cdot \vect{v}_i.
\end{equation}
This is the heat current expression implemented in LAMMPS. However, as pointed out in Ref. \cite{fan2015}, this stress-based formula does not apply to many-body potentials. The correct potential part of the heat current formula for many-body potentials reads
\begin{equation}
\vect{J} = \sum_i \sum_{j \neq i} \vect{r}_{ij}
\left(
      \frac{\partial U_j}{\partial \vect{r}_{ji}} \cdot \vect{v}_i
\right).
\end{equation}
Similar to the case of virial stress, one can also decompose the total heat current into per-atom ones:
\begin{equation}
\vect{J} = \sum_i \vect{J}_i,
\end{equation}
\begin{equation}
\vect{J}_i = \sum_{j\neq i}   \vect{r}_{ij}
\left(
      \frac{\partial U_j}{\partial \vect{r}_{ji}} \cdot \vect{v}_i
\right).
\end{equation}

\begin{algorithm}
\caption{Pseudo code for the force evaluation kernel for many-body potentials in GPUMD.}
\label{algorithm:new}
\begin{algorithmic}[1]
\Require $b$ is the block index
\Require $t$ is the thread index
\Require $S_b$ is the block size
\Require $i=S_b\times b+t$
\State Initialize $U_i$, $\vect{F}_i$, $\textbf{W}_i$, and $\vect{J}_i$ to zero
\If {$i<N$}
    \State Read in $\vect{r}_i$ from global memory
    \State Read in $\vect{v}_i$ from global memory
    \For {$m$ = 0 to NN$_i - 1$}
        \State $j \leftarrow \text{NL}_{im}$
        \State Read in $\vect{r}_j$ from global memory and calculate
               $\vect{r}_{ij}$
        \State $\vect{r}_{ij} \leftarrow $ minimum image of $\vect{r}_{ij}$
        \State Calculate $U_{ij}$
        \State Calculate $\frac{\partial U_i}{\partial \vect{r}_{ij}}$
               (with a for-loop over neighbours of $i$)
        \State Calculate $\frac{\partial U_j}{\partial \vect{r}_{ji}}$
               (with a for-loop over neighbours of $j$)
        \State $U_i \leftarrow U_i + \frac{1}{2} U_{ij}$
        \State $\vect{F}_i \leftarrow \vect{F}_i +
               \left(\frac{\partial U_i}{\partial \vect{r}_{ij}} -
               \frac{\partial U_j}{\partial \vect{r}_{ji}}\right)$
        \State $\textbf{W}_i \leftarrow \vect{W}_i
               - \frac{1}{2} \vect{r}_{ij} \otimes
               \left(\frac{\partial U_i}{\partial \vect{r}_{ij}} -
               \frac{\partial U_j}{\partial \vect{r}_{ji}}\right)$
        \State $\vect{J}_i \leftarrow \vect{J}_i
               + \vect{r}_{ij}\left(
               \frac{\partial U_j}{\partial \vect{r}_{ji}}
               \cdot \vect{v}_i \right)$
    \EndFor
    \State Save the per-atom quantities $U_i$, $\vect{F}_i$, $\textbf{W}_i$, and
           $\vect{J}_i$ to global memory
\EndIf
\end{algorithmic}
\end{algorithm}

\begin{algorithm}
\caption{Pseudo code for precomputing the bond-order functions and their derivatives for Tersoff-type potentials in GPUMD.}
\label{algorithm:bij}
\begin{algorithmic}[1]
\Require $b$ is the block index
\Require $t$ is the thread index
\Require $S_b$ is the block size
\Require $i=S_b\times b+t$
\If {$i<N$}
    \State Read in $\vect{r}_i$ from global memory
    \For {$m$ = 0 to NN$_i - 1$}
        \State $j \leftarrow \text{NL}_{im}$
        \State Read in $\vect{r}_j$ from global memory and calculate
               $\vect{r}_{ij}$
        \State $\vect{r}_{ij} \leftarrow $ minimum image of $\vect{r}_{ij}$
        \State Calculate $\zeta_{ij}$ (with a for-loop over neighbours of $i$)
        \State Calculate $b_{ij}$ and $\partial b_{ij}/\partial \zeta_{ij}$
        \State Save $b_{ij}$ and $\partial b_{ij}/\partial \zeta_{ij}$
               to global memory
    \EndFor
\EndIf
\end{algorithmic}
\end{algorithm}

From the above discussion, we see that all the relevant quantities have a simple per-atom expression. This is exactly what one needs for an efficient GPU-implementation: the per-atom quantities ($U_i$, $\vect{F}_i$, $\vect{W}_i$, and $\vect{J}_i$) can be accumulated solely by the thread associated with atom $i$ and no write conflict would occur. In Algorithm \ref{algorithm:new}, we present a pseudo code for the force evaluation kernel on the GPU. This algorithm is much simpler than Algorithm \ref{algorithm:conventional_algorithm} and can be straightforwardly implemented using CUDA or OpenCL.

There is a subtle technical point for Tersoff-type potentials. As can be seen from  Eq. (\ref{equation:F_ij_New}), the bond-order function $b_{ij}$ and their derivatives $\partial b_{ij}/\partial \zeta_{ij}$ will be frequently used in the force evaluation kernel. We can reduce the amount of redundant calculations by using a two-kernel approach, where the first kernel is used to precompute $b_{ij}$ and $\partial b_{ij}/\partial \zeta_{ij}$, and the second kernel is used to perform the force evaluation as in Algorithm \ref{algorithm:new}. The pseudo code for the first kernel is presented in Algorithm \ref{algorithm:bij}. This kernel only takes up about $20\%$ of the computation time for the whole force evaluation, but using the two-kernel approach reduces the amount of calculations for $b_{ij}$ and $\partial b_{ij} / \partial \zeta_{ij}$ by a factor of about $M$, where $M$ is the maximum number of neighbours per atom. In contrast, there is no need of precomputing for the SW potential, and only a single kernel is needed for the force evaluation.

\subsection{Generalization to other many-body potentials}

The above formalism for the Tersoff potential also applies to other many-body potentials. In Ref. \cite{fan2015}, it has been shown that for any many-body potential, the force, virial stress tensor, and heat current have the following per-atom forms:
\begin{equation}
  \vect{F}_i^{\text{many-body}} = \sum_j
    \left( \frac{\partial U_i}{\partial \vect{r}_{ij}} -
    \frac{\partial U_j}{\partial \vect{r}_{ji}} \right),
\end{equation}
\begin{equation}
  \textbf{W}_i^{\text{many-body}} = -\frac{1}{2} \sum_{j\neq i}
  \vect{r}_{ij} \otimes
  \left( \frac{\partial U_i}{\partial \vect{r}_{ij}} -
   \frac{\partial U_j}{\partial \vect{r}_{ji}} \right),
\end{equation}
\begin{equation}
  \vect{J}_i^{\text{many-body}} = \sum_{j\neq i}   \vect{r}_{ij}
\left(
      \frac{\partial U_j}{\partial \vect{r}_{ji}} \cdot \vect{v}_i
\right).
\end{equation}
Thanks to these per-atom expressions, one can construct a force evaluation CUDA kernel for any many-body potential without worrying about write conflicts.

\subsection{Comparison with previous works}

Our force evaluation algorithm should be largely equivalent to that proposed by Brown and Yamada \cite{brown2013} for the SW potential, which was recently generalized to Tersoff-type potentials by Nguyen \cite{nguyen2017}. For example, the first and last two \verb"attractive" functions in Listing 1 of Ref. \cite{nguyen2017} should be equivalent to lines 10 and 11, respectively, in our Algorithm \ref{algorithm:new}. However, we stress that (1) Our algorithm is original and general, which is based on the formalisms developed in Ref. \cite{fan2015}; (2) Useful quantities such as per-atom virial stress and heat current can also be unambiguously and efficiently calculated along with the force evaluation in our algorithm; (3) We only need a single kernel (except for Tersoff-type potentials where we intentionally precompute the bond order functions) for force evaluation, while more kernels are used in previous works \cite{brown2013,nguyen2017}.

\section{Performance evaluation}
\label{section:Performance}

\begin{figure}
\begin{center}
  \includegraphics[width=0.9\columnwidth]{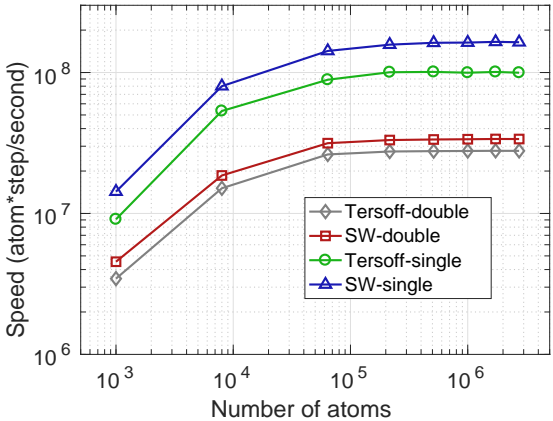}
  \caption{(color online) Computational speed of GPUMD as a function of number of atoms for Tersoff and Stillinger-Weber potentials using double- or single-precision. The test system is silicon crystal at 300 K and zero pressure. A Tesla K40 card is used to run the code.}
  \label{figure:speed}
\end{center}
\end{figure}

\begin{figure}[htb]
\begin{center}
  \includegraphics[width=0.9\columnwidth]{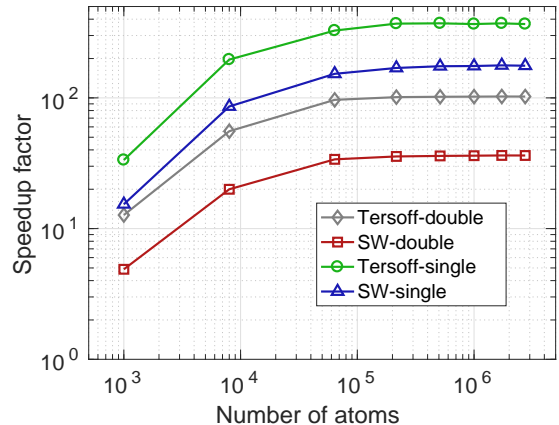}
  \caption{(color online) Speedup factor of GPUMD (running on a Tesla K40 GPU) as a function of the number of atoms with respect to the serial version of LAMMPS running on Intel Xeon CPU X5670 @ 2.93 GHz. The test system is silicon crystal at 300 K and zero pressure. }
  \label{figure:speedup}
\end{center}
\end{figure}

\begin{figure}[htb]
\begin{center}
  \includegraphics[width=0.9\columnwidth]{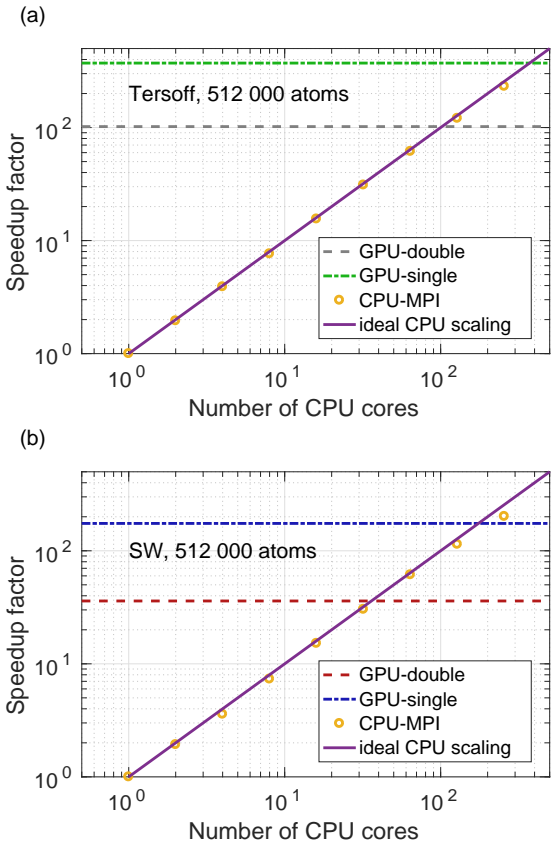}
  \caption{(color online) The markers show the scaling of the performance (relative to a single CPU core) of the MPI version of LAMMPS with respect to the number of CPU cores (Intel Xeon CPU X5670 @ 2.93 GHz). The dashed lines show the performance of GPUMD running on a Tesla K40 GPU. Panel (a) refers to the Tersoff potential and panel (b) the Stillinger-Weber potential. The solid line in each panel indicates the ideal scaling that can be achieved by MPI parallelism. The test system is silicon crystal with 512 000 atoms.}
  \label{figure:gpu_vs_many_cpus}
\end{center}
\end{figure}

\begin{figure}[htb]
\begin{center}
  \includegraphics[width=0.9\columnwidth]{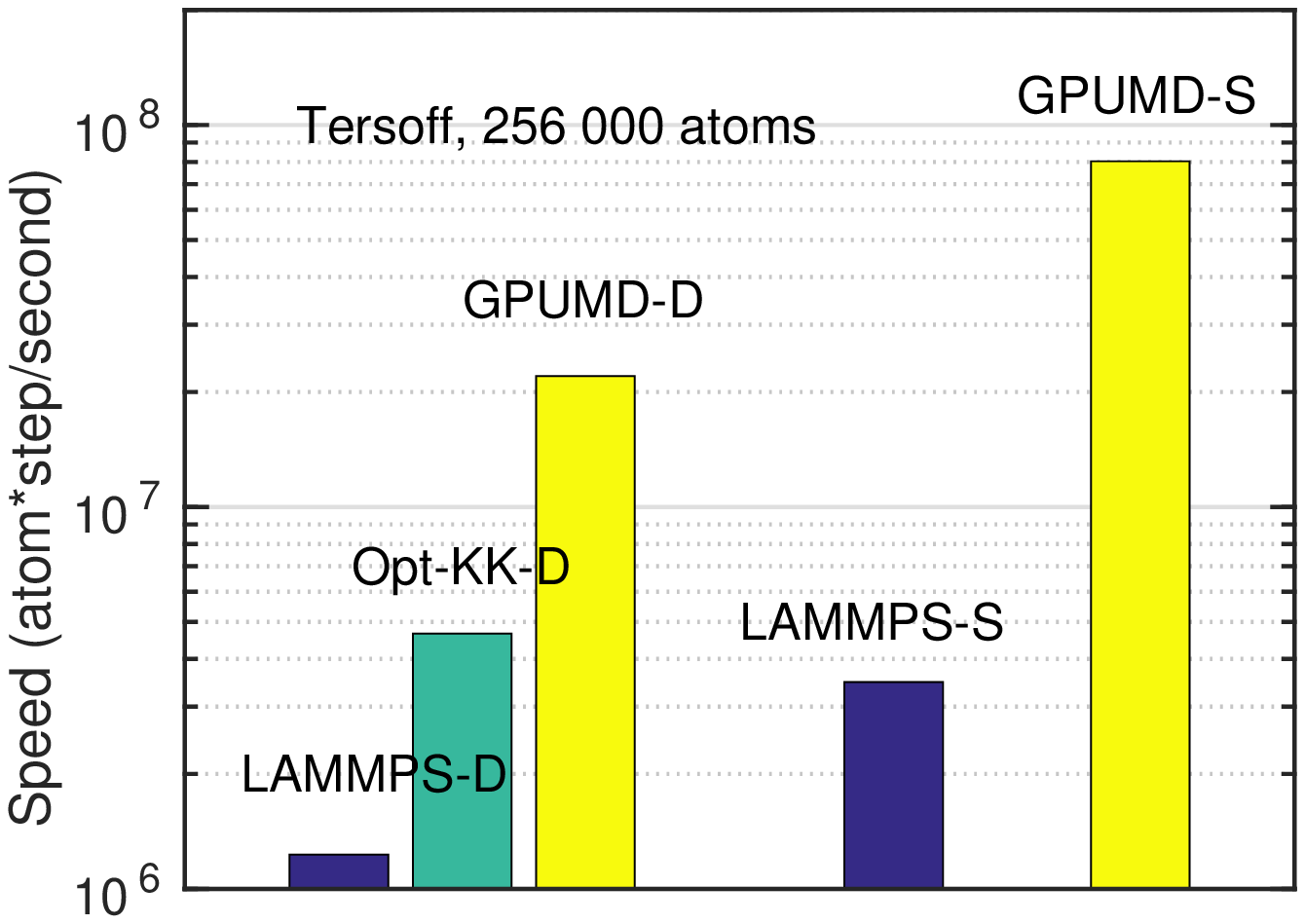}
  \caption{(color online) A comparison between the performances of GPUMD (labelled as GPUMD-D and GPUMD-S for single- and double-precision), a GPU version of LAMMPS (labelled as LAMMPS-D and LAMMPS-S for single- and double-precision), and an optimized version by H\"{o}hnerbach \textit{et al.} \cite{hohnerbach2016} based on LAMMPS (labelled as Opt-KK-D for double-precision). The test system is silicon crystal with 256 000 atoms and the Tersoff potential is used.
The data for LAMMPS-S, LAMMPS-D, and Opt-KK-D are taken from Ref. \cite{hohnerbach_github} provided by H\"{o}hnerbach \textit{et al.}. All the codes run on a Tesla K40 GPU.}
  \label{figure:gpu_vs_gpu}
\end{center}
\end{figure}

We have implemented the force evaluation kernels for the Tersoff and Stillinger-Weber potentials into our GPUMD code using CUDA, and the code has already been used to study various problems in large-scale (up to a few million atoms) graphene-based materials \cite{hirvonen2016,mortazavi2016,fan2016,fan2017prb}.  Systematic applications on thermal transport calculations have been presented in Ref. \cite{fan2015}. Here, we measure the performance of GPUMD and compare it with that of LAMMPS, which is the standard production code for simulations with many-body potentials.

Simulations with LAMMPS (version 9 Dec 2014) are performed using the ``TianHe-1A'' supercomputer. Each compute node has two CPUs, which are Intel Xeon X5670 @ 2.93GHz released in 2010. Each CPU has 6 cores, 16 GB memory, a peak performance of 70 GFLOPS, and a maximum power consumption of 95 W. Simulations with GPUMD are performed using a workstation equipped with a Tesla K40 GPU released in 2013, which has 12 GB device memory, a peak performance of 1.4 TFLOPS in double-precision arithmetic and 4.3 TFLOPS in single-precision arithmetic, and a maximum power consumption of 235 W.

The test system is silicon crystal of cubic domain with the number of atoms varying from $N = 5 \times 5 \times 5 \times 8  =1000$ to $N = 70 \times 70 \times 70 \times 8  =2~744~000$. For each domain size, we run a simulation in the $NPT$ ensemble (300 K and 0 Pa, using the Berendsen weak-coupling thermostat and barostat, and the velocity-Verlet integration approach) for 1000 steps and record the computation time $t_{\text{comp}}$ (in unit of second). The computational speed is then calculated as $N \times 1000/t_{\text{comp}}$, which has a unit of atom $\times$ step / second. A Verlet neighbour list (each atom has 4 neighbours) is constructed in the beginning and is not rebuilt during the time evolution (there is one exception, though, which will be stated later). The time-integration and the force-evaluation parts take up about $10\%$ and $90\%$ of the whole computation time, respectively.

Figure \ref{figure:speed} shows the scaling of the computational speed of GPUMD with respect to the number of atoms for the Tersoff and Stillinger-Weber potentials, with either double- or single-precision, using a Tesla K40 GPU. For all the cases, the speed increases quickly with increasing $N$ and almost saturates when $N$ exceeds $10^5$. The single-precision version for each potential is about 3-4 times as fast as the double-precision version. This is partly due to the faster single-precision floating point arithmetic, and partly due to the fact that the double-precision version uses more registers in the force-evaluation kernel and has lower GPU occupancy. Detailed profiling shows that the occupancy is about 50\% for the single-precision version and about 25\% for the double-precision version. Despite these relatively low occupancies, the computational speeds shown in Fig. \ref{figure:speed} are impressive. For example, the single-precision version of the Stillinger-Weber potential can achieve a speed of 25 ns per day (with a time step of 2 fs) for a system with $10^6$ atoms.

To better appreciate the high performance achieved by GPUMD, we compare its performance against that obtained by LAMMPS running on a single CPU core, which is Intel Xeon CPU X5670 @ 2.93 GHz. The single-core speed of LAMMPS on this CPU model is $2.7 \times 10^5$ atom $\times$ step / second for the Tersoff potential and $9.3 \times 10^5$ atom $\times$ step / second for the SW potential. The speedup factors for GPUMD running on a Tesla K40 GPU are presented in Fig. \ref{figure:speedup}. For relatively large systems, the speedup factors range from a few tens to a few hundred, depending on the types of potential and floating point arithmetic. We note that the CPU available to us is about 3 years older than the GPU used and using a CPU released at the same year as the GPU would roughly double the LAMMPS speeds and halve the speedup factors achieved by GPUMD.

To give a more realistic comparison between the performance of GPUMD and the CPU version of LAMMPS, we consider a system of 512 000 atoms and run the LAMMPS code with varying number (from 1 to 256) of CPU cores of the same specification as above using MPI parallelism. It can be seen from Fig. \ref{figure:gpu_vs_many_cpus} that the MPI version of LAMMPS scales quite well up to about 100 CPU cores, but starts to scale less ideally afterwards. Taking the Tersoff potential as an example, the computational speeds for the double- and single-precision versions of GPUMD are equivalent to those of LAMMPS with about 100 and 500 CPU cores, respectively.

Finally, Fig. \ref{figure:gpu_vs_gpu} also gives a comparison between the performance of GPUMD and those of some available GPU versions of LAMMPS (taken from Ref. \cite{hohnerbach_github}) using the Tersoff potential. Here, to be consistent with the simulations by H\"{o}hnerbach \textit{et al.} \cite{hohnerbach2016}, we update the neighbour list every 5 time steps, which reduces the performance by about 20\% compared to the case without rebuilding the neighbour list. For both double- and single-precision, GPUMD is more than one order of magnitude faster than the original version of LAMMPS. The optimized version by H\"{o}hnerbach \textit{et al.} \cite{hohnerbach2016} is a few times faster than the original version of LAMMPS, but is still a few times slower than GPUMD.

\section{Conclusions}
\label{section:Conclusion}

In summary, we analysed the difficulty in implementing many-body potentials in MD simulations on graphics processing units and presented an efficient algorithm based on an explicit pairwise force expression for many-body potentials. In this algorithm, the virial stress tensor and the heat current also have well-defined per-atom expressions. Therefore, the force, virial stress, and heat current for a given atom can be accumulated within in a single thread and the algorithm is free of write conflict by construction. This crucial property allows for a simple, flexible, and efficient implementation of any many-body potential on the GPU.

We have implemented the algorithm for the Tersoff and Stillinger-Weber potentials in our GPUMD code, which has excellent performance. GPUMD running on a single Tesla K40 GPU can be as fast as LAMMPS running with tens to hundreds of CPU cores. Our code is available upon request and will be made public later.

\section*{Acknowledgements}

This work was supported by the Academy of Finland through its Centres of Excellence Programme (2015-2017) under project number 284621 and  National Natural Science Foundation of China under Grants No. 11404033 and 11504384.  We acknowledge the computational resources provided by Aalto Science-IT project, Finland's IT Center for Science (CSC), and China Scientific Computing Grid (ScGrid). We thank the great help from the GPU experts from CSC and NVIDIA during the GPU hackathon organized by Sebastian von Alfthan.

\bibliographystyle{99}

\begin{thebibliography}{99}

\bibitem{gotz2012}
A. W. G\"{o}tz, M. J. Williamson, D. Xu, D. Poole, S. Le Grand, R. C. Walker,
Routine Microsecond Molecular Dynamics Simulations with AMBER on GPUs. 1. Generalized Born,
J. Chem. Theory Comput. 8 (2012) 1542-1555.

\bibitem{pall2013}
S. P\'{a}ll, B. Hess,
A flexible algorithm for calculating pair interactions on SIMD architectures,
Comput. Phys. Comm. 184 (2013) 2641-2650.

\bibitem{kutzner2015}
C. Kutzner, S. P\'{a}ll, M. Fechner, A. Esztermann, B. L. de Groot, H. Grubm\"{u}ller,
Best bang for your buck: GPU nodes for GROMACS biomolecular simulations,
J. Comput. Chem. 36 (2015) 1990-2008.

\bibitem{plimpton1995}
S. Plimpton,
Fast Parallel Algorithms for Short-Range Molecular Dynamics,
J. Comput. Phys. 117 (1995) 1-19.

\bibitem{brown2011}
W. M. Brown, P. Wang, S. J. Plimpton, A. N. Tharrington,  Implementing Molecular Dynamics on Hybrid High Performance Computers - Short Range Forces,
Comput. Phys. Comm. 182 (2011) 898-911.

\bibitem{brown2012}
W. M. Brown, A. Kohlmeyer, S. J. Plimpton, A. N. Tharrington,
Implementing Molecular Dynamics on Hybrid High Performance Computers - Particle-Particle Particle-Mesh. Comput. Phys. Comm. 183 (2012) 449-459.

\bibitem{stone2007}
J. E. Stone, J. C. Phillips, P. L. Freddolino, D. J. Hardy, L. G. Trabuco, K. Schulten, Accelerating molecular modeling applications with graphics
processors,
J. Comput. Chem. 28 (2007) 2618-2640.

\bibitem{eastman2010}
P. Eastman, V. S. Pande,
Efficient Nonbonded Interactions for Molecular Dynamics on a Graphics Processing Unit,
J. Comput. Chem. 31 (2010) 1268-1272.

\bibitem{anderson2008}
J. A. Anderson, C. D. Lorenz, A. Travesset,
General purpose molecular dynamics simulations fully implemented on graphics processing units,
J. Comput. Phys. 227 (2008) 5342-5359.

\bibitem{anderson2013}
J. A. Anderson, S. C. Clotzer,
The development and expansion of HOOMD-blue through six years of GPU proliferation,
arXiv:1308.5587v1.

\bibitem{glaser2015}
J. Glaser, T. D. Nguyen, J. A. Anderson, P. Lui, F. Spiga, J. A. Millan, D. C. Morse, S. C. Glotzer,
Strong scaling of general-purpose molecular dynamics simulations on GPUs,
Comput. Phys. Comm. 192 (2015) 97-107.

\bibitem{colberg2011}
P. H. Colberg, F. H\"ofling,
Highly accelerated  simulations of glassy dynamics using GPUs: Caveats on limited floating-point precision,
Comput. Phys. Comm. 182  (2011) 1120-1129.

\bibitem{bailey2015}
N. P. Bailey \textit{et al.},
RUMD: A general purpose molecular dynamics package optimized to utilize GPU hardware down to a few thousand particles,
arXiv:1506.05094v2.

\bibitem{morozov2011}
I. V. Morozov, A. M. Kazennov, R. G. Bystryi, G. E. Norman, V. V. Pisarev, V. V. Stegailov,
Molecular dynamics simulations of the relaxation processes in the condensed matter on GPUs,
Comput. Phys. Comm. 182 (2011) 1974-1978.

\bibitem{brown2012b}
W. M. Brown, T. D. Nguyen, M. Fuentes-Cabrera, J. D. Fowlkes, P. D. Rack, M. Berger, A. S. Bland,
An evaluation of Molecular Dynamics performance on the hybrid Cray XK6 supercomputer,
Procedia Computer Science 9 (2012) 186-195.

\bibitem{hou_qing2013}
Q. Hou, M. Li, Y. Zhou, J. Cui, Z. Cui, J. Wang,
Molecular dynamics simulations with many-body potentials on multiple GPUs - The implementation, package and performance,
Comput. Phys. Comm. 184 (2013) 2091-2101.

\bibitem{tersoff1989}
J. Tersoff,
Modeling solid-state chemistry: Interatomic potentials for mnlticomponent systems,
Phys. Rev. B 39 (1989) 5566-5568.


\bibitem{stillinger1985}
F. H. Stillinger and T. A. Weber,
Computer simulation of local order in condensed phases of silicon,
Phys Rev B, 31 (1985) 5262-5271.

\bibitem{brenner2002}
D. W. Brenner, O. A. Shenderova, J. A. Harrison, S. J. Stuart, B. Ni, and S. B. Sinnott,
A second-generation reactive empirical bond order (REBO) potential energy expression for hydrocarbons,
J. Phys.: Condens. Matter 14 (2002) 783–802.


\bibitem{hou2013}
C. Hou, J. Xu, P. Wang, W. Huang, and X. Wang,
Efficient GPU-accelerated molecular dynamics simulation of solid covalent crystals,
Comput. Phys. Comm. 184 (2013) 1364-1371.

\bibitem{brown2013}
W. M. Brown, M. Yamada,
Implementing Molecular Dynamics on Hybrid High Performance Computers - Three-Body Potentials,
Comput. Phys. Comm. 184  (2013) 2785-2793.

\bibitem{knizhnik2015}
A. A. Knizhnik, A. S. Minkin, B. V. Potapkin,
OpenCL realization of some many-body potentials,
Computer Research and Modeling 7 (2015) 549-558.

\bibitem{hohnerbach2016}
M. H\"ohnerhach, A. E. Ismail, P. Bientinesi,
The vectorization of the Tersoff multi-body potential: An exercise in performance portability,
arXiv:1607.02904v1.

\bibitem{tredak2016}
P. Tredak, W. R. Rudnicki, and J. A. Majewski
Efficient implementation of the many-body Reactive Bond Order (REBO) potential on GPU,
J. Comput. Phys. 321, (2016) 556–570.


\bibitem{nguyen2017}
T. D. Nguyen,
GPU-accelerated Tersoff potentials for massively parallel molecular dynamics simulations,
Comput. Phys. Comm. 212, (2017) 113-122.

\bibitem{cuda}
http://docs.nvidia.com/cuda/cuda-c-programming-guide/

\bibitem{fan2015}
Z. Fan, L. F. C. Pereira, H.-Q. Wang, J.-C. Zheng, D. Donadio, A. Harju,
Force and heat current formulas for many-body potentials in molecular dynamics simulation with applications to thermal conductivity calculations,
Phys. Rev. B 92 (2015) 094301.

\bibitem{fan2013}
Z. Fan, T. Siro, A. Harju,
Accelerated molecular dynamics force evaluation on graphics processing units for thermal conductivity calculations,
Comput. Phys. Comm. 184  (2013) 1414-1425.


\bibitem{howard2016}
M. P. Howard, J. A. Anderson, A. Nikoubashman, S. C. Glotzer, A. Z. Panagiotopoulos,
Efficient neighbour list calculation for molecular simulation of colloidal systems using graphics processing units,
Comput. Phys. Comm. 203 (2016) 45-52.


\bibitem{hirvonen2016}
P. Hirvonen, M. M. Ervasti, Z. Fan, M. Jalalvand, M. Seymour, S. M. V. Allaei, N. Provatas, A. Harju, K. R. Elder, T. Ala-Nissila,
Multiscale modeling of polycrystalline graphene: A comparison of structure and defect energies of realistic samples from phase field crystal models,
Phys. Rev. B 94 (2016) 035414.

\bibitem{mortazavi2016}
B. Mortazavi, Z. Fan, L. F. C. Pereira, A. Harju, T. Rabczuk,
Amorphized graphene: A stiff material with low thermal conductivity,
Carbon 103 (2016) 318-326.

\bibitem{fan2016}
Z. Fan, A. Uppstu, A. Harju,
Dominant source of disorder in graphene: Charged impurities or ripples?
2D Mater. 4 (2017) 025004.

\bibitem{fan2017prb}
Z. Fan, L. F. C. Pereira, P. Hirvonen, M. M. Ervasti, K. R. Elder, D. Donadio, T. Ala-Nissila, and A. Harju,
Thermal conductivity decomposition in two-dimensional materials: Application to graphene,
Phys. Rev. B 95 (2017), 144309.


\bibitem{hohnerbach_github}
http://github.com/HPAC/lammps-tersoff-vector.

\end{thebibliography}

\end{document}